\documentclass[acus]{pac99}
\usepackage{epsfig}
\newcommand{\ud}{\mathrm{d}}
\begin{document}
\title{NONLINEAR ACCELERATOR PROBLEMS VIA WAVELETS:\\
4. SPIN-ORBITAL MOTION}
\author{A.~Fedorova, M.~Zeitlin, IPME, RAS, St.~Petersburg, Russia
\thanks{e-mail: zeitlin@math.ipme.ru}
\thanks{http://www.ipme.ru/zeitlin.html;
        http://www.ipme.nw.ru/zeitlin.html}}
\maketitle
\begin{abstract}
In this series of eight papers  we
present the applications of methods from
wavelet analysis to polynomial approximations for
a number of accelerator physics problems.
In this part we consider a model for spin-orbital motion:
 orbital
dynamics and Thomas-BMT equations for classical spin vector. We
represent the solution of this dynamical system
 in framework of biorthogonal wavelets via variational
approach.
We consider a different variational approach, which is
applied to each scale.
\end{abstract}
\section{INTRODUCTION}
This is the fourth part of our eight presentations in which we consider
applications of methods from wavelet analysis to nonlinear accelerator
physics problems. This is a continuation of our results from [1]-[8],
which is based on our approach  to investigation
of nonlinear problems -- general, with additional structures (Hamiltonian,
symplectic or quasicomplex), chaotic, quasiclassical, quantum, which are
considered in the framework of local (nonlinear) Fourier analysis, or wavelet
analysis. Wavelet analysis is a relatively novel set of mathematical
methods, which gives us a possibility to work with well-localized bases in
functional spaces and with the general type of operators (differential,
integral, pseudodifferential) in such bases.
In  this part we consider
spin orbital motion.
In section 3 we consider generalization of our approach
from part 1 to variational formulation in the biorthogonal bases of compactly
supported wavelets.
In section 4 we consider the different variational
multiresolution approach which gives  us possibility
for computations in each scale separately.

\section{Spin-Orbital Motion}
Let us consider the system of equations for orbital motion
and Thomas-BMT equation for classical spin vector [9]:
$
\ud q/\ud t={\partial H_{orb}}/{\partial p}, \quad
{\ud p}/{\ud t}=-{\partial H_{orb}}/{\partial q}
$, $\quad\ud s/\ud t=w\times s$,
where
\begin{eqnarray}
H_{orb}&=&c\sqrt{\pi^2+m_0c^2}+e\Phi,\nonumber\\
w=&-&\frac{e}{m_0 c \gamma} (1+\gamma G)\vec B\\
  &+&\frac{e}{m_0^3 c^3\gamma}\frac{G(\vec\pi\cdot\vec B)
\vec\pi}{(1+\gamma)}\nonumber\\
 &+&\frac{e}{m_0^2 c^2\gamma}\frac{G +\gamma G+1}{(1+\gamma)}
[\pi\times E],\nonumber
\end{eqnarray}
$q=(q_1,q_2,q_3), p=(p_1,p_2,p_3)$ are canonical position and momentum,
$s=(s_1,s_2,s_3)$ is the classical spin vector of length $\hbar/2$,
$\pi=(\pi_1,\pi_2,\pi_3)$ is kinetic momentum vector.
We may introduce in 9-dimensional phase space $z=(q,p,s)$ the Poisson brackets
$
\{f(z),g(z)\}=f_qg_p-f_pg_q+[f_s\times g_s]\cdot s
$
and the  Hamiltonian equations are
$
{\ud z}/{\ud t}=\{z,H\}
$
with Hamiltonian
\begin{equation}
H=H_{orb}(q,p,t)+w(q,p,t)\cdot s.
\end{equation}
More explicitly we have
\begin{eqnarray}
\frac{\ud q}{\ud t}&=&\frac{\partial H_{orb}}{\partial p}+\frac{\partial(w\cdot
  s)}{\partial p}\nonumber\\
\frac{\ud p}{\ud t}&=&-\frac{\partial H_{orb}}{\partial q}-\frac{\partial(w\cdot
  s)}{\partial q}\\
\frac{\ud s}{\ud t}&=&[w\times s]\nonumber
\end{eqnarray}
We will consider this dynamical system also in  another paper
via invariant approach, based on consideration of Lie-Poison structures on
semidirect products.
But from the point of view which we used in this paper we may consider the
similar approximations as in the preceding parts and then we also arrive to
some type of polynomial dynamics.

\section{VARIATIONAL APPROACH IN BIORTHO\-GONAL WAVELET BASES}

Because integrand of variational functionals is represented
by bilinear form (scalar product) it seems more reasonable to
consider wavelet constructions [10] which take into account all advantages of
this structure.
The action functional for loops in the phase space is [11]
\begin{equation}
F(\gamma)=\displaystyle\int_\gamma pdq-\int_0^1H(t,\gamma(t))dt
\end{equation}
The critical points of $F$ are those loops $\gamma$, which solve
the Hamiltonian equations associated with the Hamiltonian $H$
and hence are periodic orbits. By the way, all critical points of $F$ are
the saddle points of infinite Morse index, but surprisingly this approach  is
very effective. This will be demonstrated using several
variational techniques starting from minimax due to Rabinowitz
and ending with Floer homology. So, $(M,\omega)$ is symplectic
manifolds, $H: M \to R $, $H$ is Hamiltonian, $X_H$ is
unique Hamiltonian vector field defined  by
$
\omega(X_H(x),\upsilon)=-dH(x)(\upsilon),\quad \upsilon\in T_xM,
\quad x\in M,
$
where $ \omega$ is the symplectic structure.
A T-periodic solution $x(t)$ of the Hamiltonian equations
$
\dot x=X_H(x)
$ on M
is a solution, satisfying the boundary conditions $x(T)$ $=x(0), T>0$.
Let us consider the loop space $\Omega=C^\infty(S^1, R^{2n})$,
where $S^1=R/{\bf Z}$, of smooth loops in $R^{2n}$.
Let us define a function $\Phi: \Omega\to R $ by setting
\begin{equation}
\Phi(x)=\displaystyle\int_0^1\frac{1}{2}<-J\dot x, x>dt-
\int_0^1 H(x(t))dt, \quad x\in\Omega
\end{equation}
The critical points of $\Phi$ are the periodic solutions of $\dot x=X_H(x)$.
Computing the derivative at $x\in\Omega$ in the direction of $y\in\Omega$,
we find
\begin{eqnarray}
&&\Phi'(x)(y)=\frac{d}{d\epsilon}\Phi(x+\epsilon y)\vert_{\epsilon=0}
=\\
&&\displaystyle\int_0^1<-J\dot x-\bigtriangledown H(x),y>dt\nonumber
\end{eqnarray}
Consequently, $\Phi'(x)(y)=0$ for all $y\in\Omega$ iff the loop $x$ satisfies
the equation
\begin{equation}
-J\dot x(t)-\bigtriangledown H(x(t))=0,
\end{equation}
i.e. $x(t)$ is a solution of the Hamiltonian equations, which also satisfies
$x(0)=x(1)$, i.e. periodic of period 1. Periodic loops may be represented by
their Fourier series:
$x(t)=\sum e^{k2\pi Jt}x_k, \ x_k\in R^{2k}$,
where $J$ is quasicomplex structure. We give relations between
quasicomplex structure and wavelets in our other paper.
But now we
need to take into account underlying bilinear structure via wavelets.
We started with two hierarchical sequences of approximations spaces [10]:
\begin{eqnarray}
&&\dots V_{-2}\subset V_{-1}\subset V_{0}\subset V_{1}\subset V_{2}\dots,\\
&&\dots \widetilde{V}_{-2}\subset\widetilde{V}_{-1}\subset
\widetilde{V}_{0}\subset\widetilde{V}_{1}\subset\widetilde{V}_{2}\dots,\nonumber
\end{eqnarray}
and as usually,
$W_0$ is complement to $V_0$ in $V_1$, but now not necessarily orthogonal
complement.
New orthogonality conditions have now the following form:
\begin{equation}
\widetilde {W}_{0}\perp V_0,\quad  W_{0}\perp\widetilde{V}_{0},\quad
V_j\perp\widetilde{W}_j, \quad \widetilde{V}_j\perp W_j
\end{equation}
translates of $\psi$ $\mathrm{span}$ $ W_0$,
translates of $\tilde\psi \quad \mathrm{span} \quad\widetilde{W}_0$.
Biorthogonality conditions are
\begin{equation}
<\psi_{jk},\tilde{\psi}_{j'k'}>=
\int^\infty_{-\infty}\psi_{jk}(x)\tilde\psi_{j'k'}(x)\ud x=
\delta_{kk'}\delta_{jj'},
\end{equation}
 where
$\psi_{jk}(x)=2^{j/2}\psi(2^jx-k)$.
Functions $\varphi(x), \tilde\varphi(x-k)$ form  dual pair:
$
<\varphi(x-k),\tilde\varphi(x-\ell)>=\delta_{kl}$,
 $<\varphi(x-k),\tilde\psi(x-\ell)>=0$.
Functions $\varphi, \tilde\varphi$ generate a multiresolution analysis.
$\varphi(x-k)$, $\psi(x-k)$ are synthesis functions,
$\tilde\varphi(x-\ell)$, $\tilde\psi(x-\ell)$ are analysis functions.
Synthesis functions are biorthogonal to analysis functions. Scaling spaces
are orthogonal to dual wavelet spaces.
Two multiresolutions are intertwining
$
V_j+W_j=V_{j+1}, \quad \widetilde V_j+ \widetilde W_j = \widetilde V_{j+1}
$.
These are direct sums but not orthogonal sums.

So, our representation for solution has now the form
\begin{equation}
f(t)=\sum_{j,k}\tilde b_{jk}\psi_{jk}(t),
\end{equation}
where synthesis wavelets are used to synthesize the function. But
$\tilde b_{jk}$ come from inner products with analysis wavelets.
Biorthogonality yields
\begin{equation}
\tilde b_{\ell m}=\int f(t)\tilde{\psi}_{\ell m}(t) \ud t.
\end{equation}
So, now we can introduce this more complicated construction into
our variational approach. We have modification only on the level of
computing coefficients of reduced nonlinear algebraical system.
This new construction is more flexible.
Biorthogonal point of view is more stable under the action of large
class of operators while orthogonal (one scale for multiresolution)
is fragile, all computations are much more simpler and we accelerate
the rate of convergence. In all types of Hamiltonian calculation,
which are based on some bilinear structures (symplectic or
Poissonian structures, bilinear form of integrand in variational
integral) this framework leads to greater success.
In particular cases we may use very useful wavelet packets from Fig.~1.
\begin{figure}[ht]
\centering
\epsfig{file=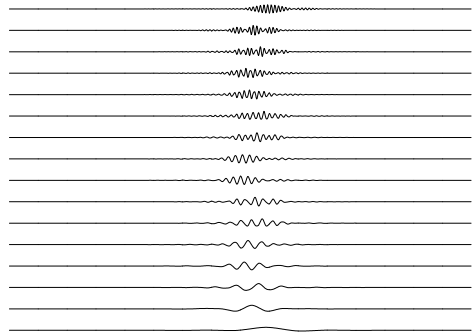, width=82.5mm, bb= 0 200 599 600, clip}
\caption{Wavelet packets.}
\end{figure}

\section{Evaluation of Nonlinearities Scale by Scale.Non-regular approximation.}
We use wavelet function $\psi(x)
$, which has $k$ vanishing moments
$
\int x^k \psi(x)\ud x=0$, or equivalently
$x^k=\sum c_\ell\varphi_\ell(x)$ for each $k$,
$0\leq k\leq K$.
Let $P_j$ be orthogonal projector on space $V_j$. By tree algorithm we
have for any $u\in L^2({\bf R})$ and $\ell\in{\bf Z}$, that the wavelet
coefficients of $P_\ell(u)$, i.e. the set $\{<u,\psi_{j,k}>, j\leq\ell-1,
k\in{\bf Z}\}$ can be compute using hierarchic algorithms from the set of
scaling coefficients in $V_\ell$, i.e. the set $\{<u,\varphi_{\ell,k}>,
k\in{\bf Z}\}$ [12]. Because for scaling function $\varphi$ we have in general
only $\int\varphi(x)\ud x=1$, therefore we have for any function $u\in L^2({\bf
R})$:
\begin{equation}
\lim_{j\to\infty, k2^{-j}\to x} \mid 2^{j/2}<u,\varphi_{j,k}>-u(x)\mid=0
\end{equation}
If the integer $n(\varphi)$ is the largest one such that
\begin{equation}
\int x^\alpha \varphi(x)\ud x=0 \qquad \mbox{for}\qquad 1\leq\alpha\leq n
\end{equation}
then if $u\in C^{(n+1)}$ with $u^{(n+1)}$ bounded we have for $j\to\infty$
uniformly in k:
\begin{equation}
\mid 2^{j/2}<u,\varphi_{j,k}>-u(k2^{-j})\mid=O(2^{-j(n+1)}).
\end{equation}
Such scaling functions with zero moments are very useful for us from the point
of view of time-frequency localization, because we have for Fourier component
$\hat\Phi(\omega)$ of them, that exists some $C(\varphi)\in{\bf R}$, such
that for $\omega\to0$ $\quad\hat\Phi(\omega)=1+C(\varphi)$ $\mid\omega\mid^{2r+2}$
(remember, that we consider r-regular multiresolution analysis).
Using such type of scaling functions lead to superconvergence properties for
general Galerkin approximation [12].
Now we need some estimates in each scale for non-linear terms of type $u\mapsto
f(u)=f\circ u$, where f is $C^\infty$ (in previous and future parts we consider
only truncated Taylor series action). Let us consider non regular space of
approximation $\widetilde V$ of the form
\begin{equation}\label{eq:til1}
\widetilde V=V_q\oplus\sum_{q\leq j\leq p-1} \widetilde{W_j},
\end{equation}
with $\widetilde{W_j}\subset W_j$. We need efficient and precise estimate of
$f\circ u$ on $\widetilde V$. Let us set for $q\in{\bf Z}$ and $u\in L^2({\bf
R})$
\begin{equation}
\prod f_q (u)=2^{-q/2}\sum_{k\in{\bf
Z}}f(2^{q/2}<u,\varphi_{q,k}>)\cdot\varphi_{q,k}
\end{equation}
We have the following important for us estimation (uniformly in q) for $u,
f(u)\in H^{(n+1)}$ [12]:
\begin{equation}\label{eq:til2}
\|P_q\left(f(u)\right)-\prod f_q(u)\|_{L^2}=O\left(2^{-(n+1)q}\right)
\end{equation}
For non regular spaces (\ref{eq:til1}) we set
\begin{equation}
\prod f_{\widetilde V}(u)=\prod f_q(u)+\sum_{\ell=q,p-1} P_{\widetilde {W_j}}
\prod f_{\ell+1}(u)
\end{equation}
Then we have the following estimate:
\begin{equation}
\Vert P_{\widetilde V}\left(f(u)\right)-\prod f_{\widetilde V}(u)\Vert_{L^2}=O(2^{-(n+1)q})
\end{equation}
uniformly in q and $\widetilde V$ (\ref{eq:til1}).
This estimate depends on q, not p, i.e. on the scale of the coarse grid, not on
the finest grid used in definition of $\widetilde V$. We have for total error
\begin{eqnarray}
&&\Vert f(u)-\prod f_{\widetilde V}(u)\Vert=
\Vert f(u)-P_{\widetilde V}(f(u))\Vert_{L^2}\nonumber\\
&&+\Vert P_{\widetilde V}(f(u)-\prod f_{\widetilde V}(u))\Vert_{L^2}
\end{eqnarray}
and since the projection error in $\widetilde V$:
$
\Vert f(u)-P_{\bar{V}}\left(f(u)\right)\Vert_{L^2}
$
 is much smaller than the projection error in
$V_q$ we have the improvement (20) of (\ref{eq:til2}).
In  concrete calculations and estimates it is very useful to consider
approximations in the particular case of
c-structured space:
\begin{eqnarray}
\widetilde{V}=&&V_q+\sum^{p-1}_{j=q}span\{\psi_{j,k},\\
&& k\in[2^{(j-1)}-c,
2^{(j-1)}+c]
\quad\mbox{\rm mod}\quad 2^j\}\nonumber
\end{eqnarray}

We are very grateful to M.~Cornacchia (SLAC), W.~Herrmannsfeldt (SLAC),
Mrs. J.~Kono (LBL) and
M.~Laraneta (UCLA) for their permanent encouragement.

 \end{document}